\documentstyle[psfig,12pt,aaspp4,epsf]{article}

\def\deg{\hbox{$^\circ$}}

\def\fmag{\hbox{$.\!\!^m$}}

\begin{document}

\slugcomment{\it 1999, PASP (in press)}

\title{SPECTROSCOPY OF STELLAR-LIKE OBJECTS CONTAINED IN THE SECOND BYURAKAN SURVEY. I}

\author{J.A. Stepanian}
\affil{Special Astrophysical Observatory RAS,\ Nizhnij Arkhys,
Karachai-Cherkessia, 357147 \\ Russia \\
Electronic mail: jstep@sao.ru}

\author{V.H. Chavushyan\footnotemark
\footnotetext{Visiting Astronomer, Special Astrophysical Observatory, Russian
Academy of Science},
L. Carrasco\footnotemark
\footnotetext{Also OAN, UNAM, Ensenada, M\'exico},
and H.M. Tovmassian}
\affil{Instituto Nacional de Astrof\'{\i}sica Optica y Electr\'onica,
Apartado Postal 51 y 216, C.P.72000,
Puebla, Pue., M\'exico \\
Electronic mail: vahram@inaoep.mx, carrasco@inaoep.mx and hrant@inaoep.mx}
\author{L.K. Erastova}
\affil{Byurakan Astrophysical Observatory, Byurakan, 378433 Armenia\\
Electronic mail: lke@bao.sci.am}

\begin{abstract}

The results of spectroscopic observations of 363 star-like objects from the
Second Byurakan Survey (SBS) are reported. This SBS's subsample has  proven 
to be a rich source of newly identified quasars, Seyfert type galaxies, 
degenerate stars and hot subdwarfs. In the subsample here studied, we identified 
35 new QSOs, 142 White Dwarfs (WDs) the majority of which, 114 are of DA type, 55 
subdwarfs (29 of which are sdB-type stars), 10 HBB, 16 NHB, 54 G-type and 25 F-type 
stars, two objects with composite spectra, four Cataclismic Variables (CV), two
peculiar emission line stars, 17 objects with continuous spectra, as well
as one planetary nebula. Among the 35 QSOs we have found two Broad
Absorption Line (BAL) QSOs, namely SBS~1423+500 and SBS~1435+500A. Magnitudes, 
redshifts, and slit spectra for all QSOs, also some typical spectra
of the peculiar stars are presented. We estimate the minimum surface density
of bright QSOs in redshift range $0.3<z<2.2$ to be 0.05 per sq. deg. for 
$B<17\fmag0$ and 0.10 per sq. deg. for $B<17\fmag5$. 

\end{abstract}


\section{INTRODUCTION}

   In the last two decades, several surveys have been undertaken primarily aimed
to the detection of large samples of QSOs. A variety of techniques have been
applied in attempts to select the small fraction of quasars among the 
samples of blue objects largely comprised of foreground stars with spectra that
resemble, at a given spectral resolution, those of quasars.
The objective prism surveys have been one of the most efficient selection methods applied. The SBS is one of such surveys, it was carried out
for a defined area of the sky, in which both stellar and non-stellar objects were systematically selected.

The Second Byurakan Survey (SBS) is a low resolution objective prism
survey with a limiting magnitude of $B\sim19\fmag5$. The survey covers
1000 square degrees contained within the limits defined by $07^{h} 40^{m} < \alpha <17^{h} 15^{m}$, in right ascension and by
$+49\deg < \delta < +61\deg$ in declination. The observing technique and the selection criteria for the SBS objects have been described by Markarian \&
Stepanian (1983), and Stepanian (1994).
Worth mentioning is that the selection criteria of the SBS include the presence of a strong UV
continua, emission lines, and/or peculiar energy distributions as inferred from
the objective prism spectra.
   These criteria have been successful in selecting objects
such as UVX galaxies (Markarian galaxies), as well as, a broad class of QSOs.
As a by product, a large number of peculiar stars, WDs, composite
and emission objects, hot subdwarfs and other types of objects are detected. The SBS catalog contains nearly 3500 objects, of which $\sim1700$
are galaxies, and $\sim1800$ are star-like objects.

One of the main goals of the SBS survey is to search and compile a complete
sample of bright QSOs in the magnitude range $15\fmag5<B<18\fmag0$. Such
sample is essential for determining the QSO's surface and space densities.
It is important also for understanding the origin and evolution of
quasars, for solving problems related to studies of the internal
structure of QSOs. It is as well important for probing the early Universe
through the analysis of absorption line systems formed along the line of sight.
The latter
requiring high spectral resolution, which is attainable only for the brighter QSOs.

During the last two decades in parallel with the SBS original survey we
have been carrying out follow-up spectroscopic observations of selected
objects from the SBS sample and still continue with it on the 6 m telescope of
the Special Astrophysical Observatory (SAO) in Russia. Observations of a
few hundred objects have been carried out with the MMT (USA), with the 2.6m telescope
of the Byurakan Observatory (Armenia), and with the 2.1 m telescope of the GHO
(Cananea, Mexico).
Historically, objective prism and color selected surveys have provided samples that help to answer some fundamental problems related to peculiar stars such as formation rates, space densities, lifetime, luminosity functions, etc. In this respect, the compilation of a data base of different classes of highly evolved stars at high galactic latitudes in a complete survey, that extends to fainter magnitudes and therefore, to larger distances, is very important.

 So far, we have obtained slit spectra for about 850  SBS stellar-like objects.
The data for nearly $\sim270$ QSOs, $\sim100$ Sy galaxies and $\sim100$
peculiar stars were published by  Markarian et al.(1980 -- 1987), Stepanian et al. (1990 -- 1993) and Stepanian (1994).
  
In the present paper we report the results of spectroscopic
observations of 363 relatively bright stellar objects from the SBS sample.
The observational data were obtained at the Special Astrophysical
Observatory (Russia) and Guillermo Haro Observatory (GHO) of the INAOE
in Cananea (Mexico).The $BV$ CCD photometric data for
confirmed QSOs were obtained with the 1 m telescope at SAO (Chavushyan et al.
1995, 1999).

\section{OBSERVATIONS AND DATA REDUCTION}

  Our spectroscopic observations have been carried out with the 6 m telescope of the SAO (Russia) and
with the 2.1 m telescope of the GHO at Cananea, M\'exico, during the period of 1978-1997.

Spectral observations with the SAO 6 m telescope have been carried
out since 1978. The Universal Astrophysical Grating Spectrograph (UAGS) in combination with an image tube
UM-92 has been used in the first series of observations.
The dispersion of the spectrograph was 90-100 \AA/mm with a spectral resolution
of 5-8 \AA. Since 1984 the observations were carried out with the SP-124
spectrograph equipped with a 1024-channel photon counting system (IPCS)
scanner (Drabek et al. 1985), installed at the Nasmith I focus. Later on the
Long Slit Spectrograph (LSS) equipped with a $530\times580$ pixel CCD
(Afanasiev at al. 1995), installed at the prime focus, was used. The adopted
slit width was 2 arcsec with an effective instrumental spectral resolution
of about 12 \AA\ for the IPCS and about 15 \AA\ for the LSS. The
 wavelength range covered was that of 3400 to 7100 \AA\ .  The data reduction
procedures -- cosmic ray hits removal, bias and flat field corrections,
wavelength linearization and flux calibration -- were carried out with
the SAO standard method of IPCS data reduction (Afanasiev at al. 1991)
and with the CCD data reduction software packages developed at the SAO
(Vlasyuk, 1993).

 Observation with the 2.1 m GHO telescope were carried out with the Landessternwarte
Faint Object Spetrograph (LFOSC) Zickgraff et al. (1997),
installed in the Cassegrain focus. This instrument is equipped
with a $600\times400$ pixel CCD. The adopted slit width was 2 arcsec, with the
effective instrumental spectral resolution of about 11 \AA\ in the wavelength
range form  4000 \AA\ to 7000 \AA\ . The IRAF reduction packages were used for  
data reduction and flux calibration.

\section{THE RESULTS}

  The journal of  observations is presented in Table 1. The following
information is listed in consecutive columns: 1 -- the SBS designation, 2-3 -- the coordinates
for the 1950.0 epoch measured by Bicay et al.(1999) with an accuracy of about $\sim1$ arcsec, 4 -- an
eye estimated $m_{pg}$ magnitude, given with an accuracy of about $\pm0\fmag5$ (Stepanian, 1994).
For some objects the photometric B magnitudes with two decimals
are given, those should have a probable error of $0\fmag05$, (Chavushyan et al. 1995, 1999), 5 -- the spectral type, 6 -- date of observation,
7 --  exposure time, 8 --  instrument used, 9 -- an alternative
designation of the object when available.

The numbers in the present subsample of SBS stellar objects of different spectral types are given in Table 2.

In classifying the detected white dwarfs and subdwarfs, we have adopted the
classification scheme developed by Sion et al. (1983), Green et al. (1986)
and Berg et al. (1992). The spectral characteristics associated with the
types adopted in our paper are listed by Berg et al. (1992). Furthermore, in order to achieve a uniform classification scheme for the objects repoted here, we have obtained slit spectra of known objects contained in some previous studies such as PHL (Palomar-Haro-Luyten), PG (Palomar-Green), LB (Luyten-Blue), and also of some other bright objects with previously well determined spectral types which are contained in the SBS. We suspect that the subsample of objects classified here as "Continuum spectrum" (Cont) is composed by a mixture of both BL Lac objects and DC stars.

Table 3 lists the observed emission lines and redshifts for the 35 new QSOs.
The redshifts were determined from the strongest emission lines. The mark ":"
refers to the ambiguous emission line determination. The redshifts of the
detected quasars fall in the interval defined by $z=0.1$ and $z=2.32$.

Plots of spectra of the QSOs and of some  typical high latitude blue stars are
presented in Figures 1 to 4. For the sake of homogeneity, we plotted there relative
intensities versus wavelength.

\section{CONCLUSIONS}

The subsample of stellar-like objects contained in the SBS catalogue has shown to
be a rich source of new QSOs, degenerate stars and hot subdwarfs. In the
present paper, the spectral classification of 363 studied objects was made. Thirty
five new QSOs, 142 WD most of which (114) are DA white dwarf stars, 55 subdwarfs (of which 29 are
sdB-type), 10 HBB, 16 NHB, 54 G and 25 F-type stars, two objects
(SBS~0834+576 and SBS~1309+544) have composite spectra, four cataclismic
variables (CV), two peculiar emission line stars, 17 objects with continuous
spectra, as well as one planetary nebula were identified. A detail
investigation of the latter object is under way.

Worth mentioning is that two of the detected QSOs, SBS~1423+500 and SBS~1435+500A, are
clearly BAL QSOs. While SBS 1201+517 may also be a BAL QSO, since a broad absorption
component of MgII ($\lambda$ 2798 \AA) emission line in its spectrum is suspected. Two
very strong absorption lines at $\lambda$ 4463 \AA\ (perhaps CIV in self absorption)
and at $\lambda$ 4015 \AA\ are evident in the spectrum of SBS~0937+503.

From visual inspection of the POSS, six stars in our sample, namely -- SBS~0958+532 (DAF:), SBS~1017+533 (CV), SBS1040+493~ (sd),
SBS~1050+582 (DA), SBS~1103+586 (DA) and SBS~1300+523 (DAB) have very close
companions and are probably binary systems.

The vast majority of star-like objects selected in SBS survey have turned out to be WDs, subdwarfs
and F and G type stars. The most numerous subset of stars found are DA WDs.

  Thus far, the spectroscopy of the subsample of stellar-like objects contained in the SBS, has
confirmed the nature of more than 300 new QSOs, one third of which are brighter than $B<17\fmag5$. This allow us to estimate
a absolute lower limit to the cumulative surface density of bright QSOs in the most complete range of redshifts ie. $0.3<z<2.2$. The most reliable values for the lower limit of bright QSOs in redshift range $0.3<z<2.2$ corresponds to 0.05 per sq. deg. for $B<17\fmag0$, and 0.10 per sq. deg. for $B<17\fmag5$.

The detailed analysis of the surface and spatial densities of QSOs, and of
different types of selected peculiar stars, will be carried out once we finish the 
spectroscopy of the entire sample of star-like objects contained in the SBS catalogue. 

\acknowledgements

We are grateful to V. Vlasyuk, A. Burenkov  for allowing us to use their
data analysis package and for assisting us at the 6 m telescope. We thank J.R.
Vald\'es for helping us with observations and 
J. Miram\'on, G. Miram\'on, R. Gonz\'ales and S. Noriega for excellent assistance
and technical support at the 2.1m GHO telescope. Finally, we thank N. Serafimovich for help in
preparing the present paper. This work has been partially supported by CONACYT research grants
No. 28499-E, 211290-5-1430PE, and  211290-5-0009PE. JAS, VHCh, LKE were partially supported
by the research grants No. 97-02-17168 and 1.2.2.2 from the Russian Foundation for Basic Research
and from the State Programm "Astronomy" respectively.

\clearpage

\begin{figure}[t]
\setcounter{figure}{0}
\caption{Plots of the spectra of QSOs observed with the 6 m telescope.
The vertical axis represents the relative flux the absissae represents the wavelength in \AA }
\epsfxsize=15cm\epsfbox{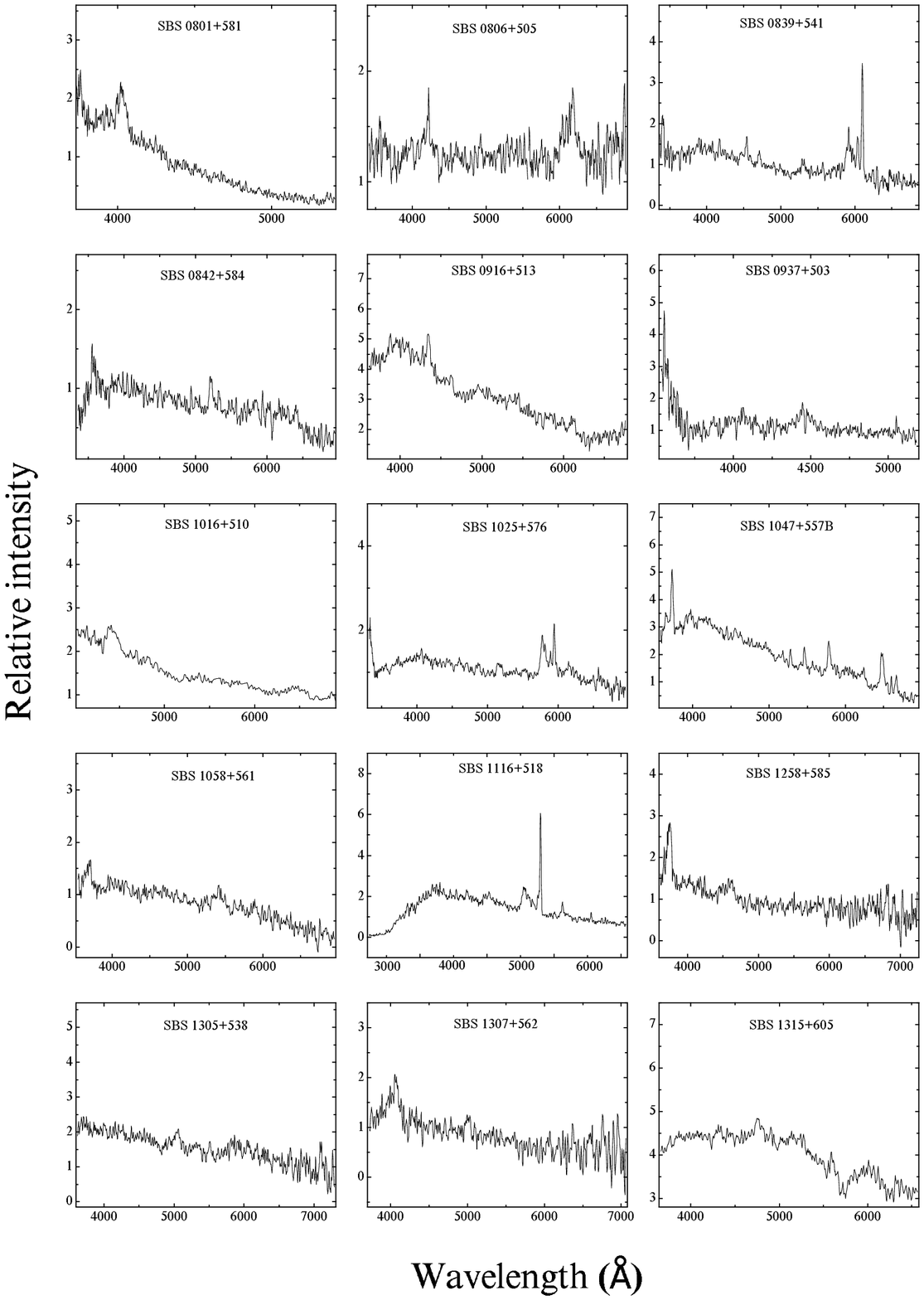}
\end{figure}

\begin{figure}[htb]
\setcounter{figure}{0}
\caption{Continued}
\epsfxsize=15cm\epsfbox{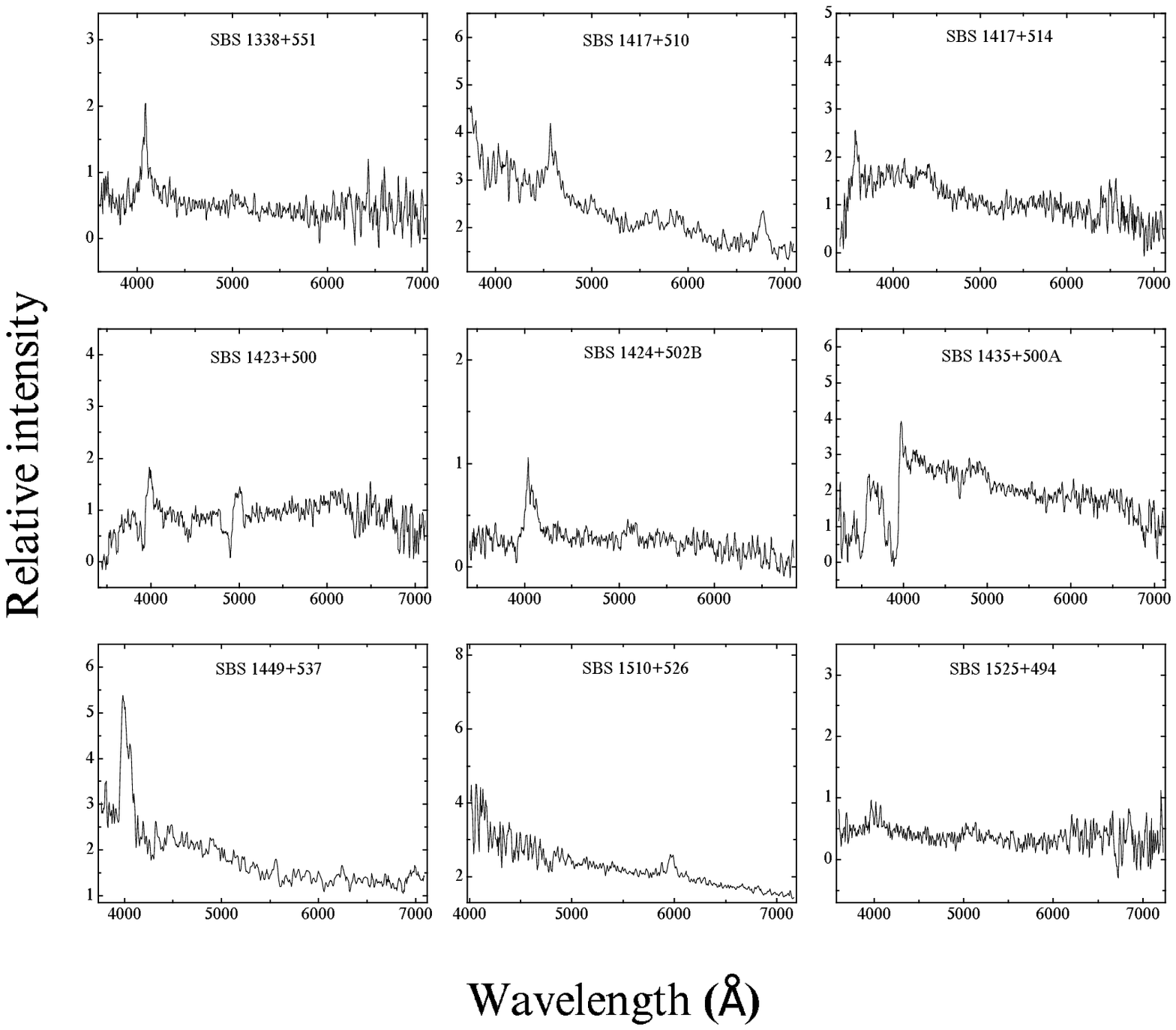}
\end{figure}

\begin{figure}[htb]
\setcounter{figure}{1}
\caption{Plots of the spectra of stars observed with the 6 m telescope.The vertical axis represents the relative flux the absissae represents the wavelength in \AA }

\epsfxsize=15cm\epsfbox{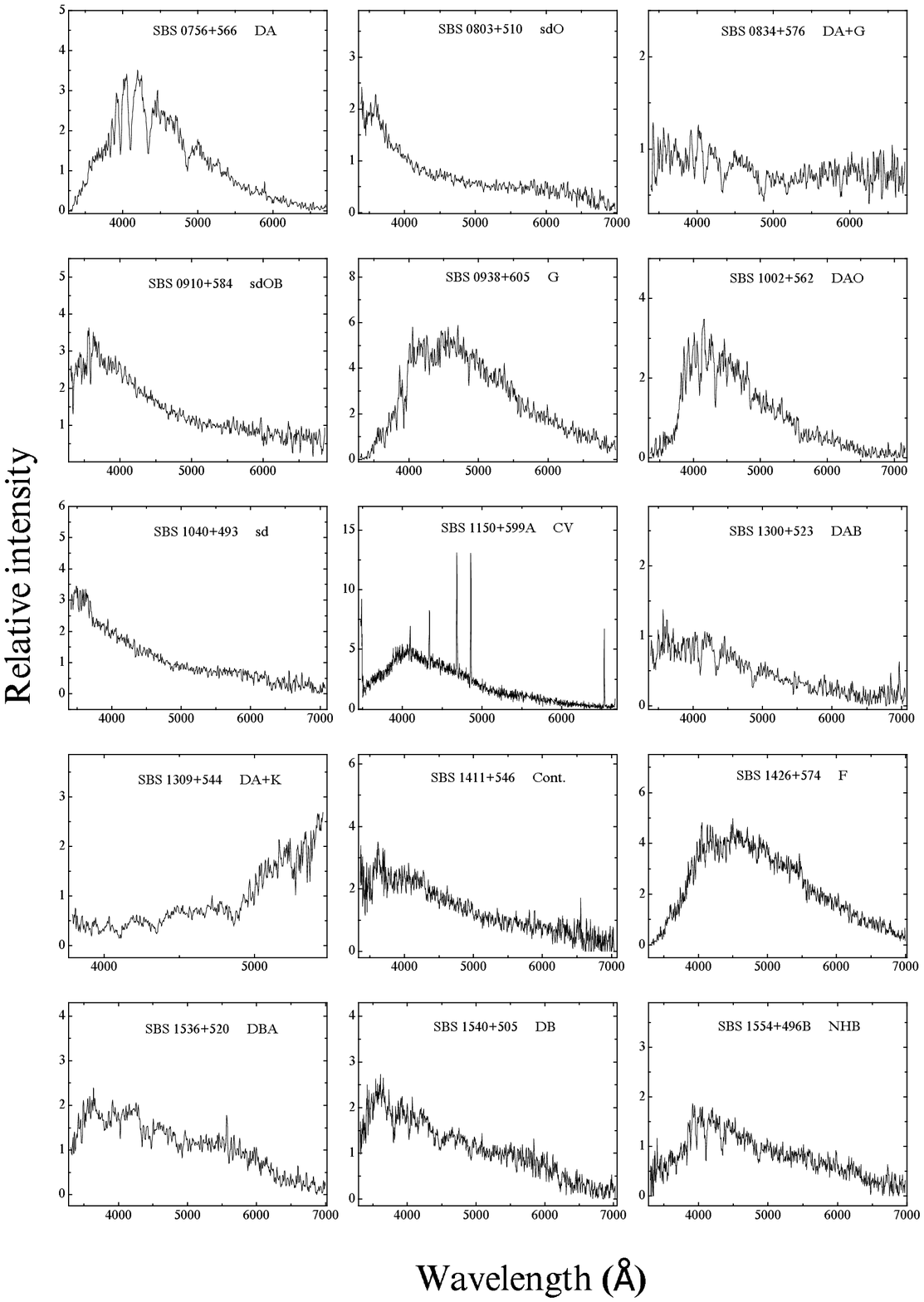}
\end{figure}

\begin{figure}[htb]
\setcounter{figure}{2}
\caption{Plots of the spectra of QSOs observed with the 2.1 m telescope.The vertical axis represents the relative flux the absissae represents the wavelength in \AA }

\epsfxsize=15cm\epsfbox{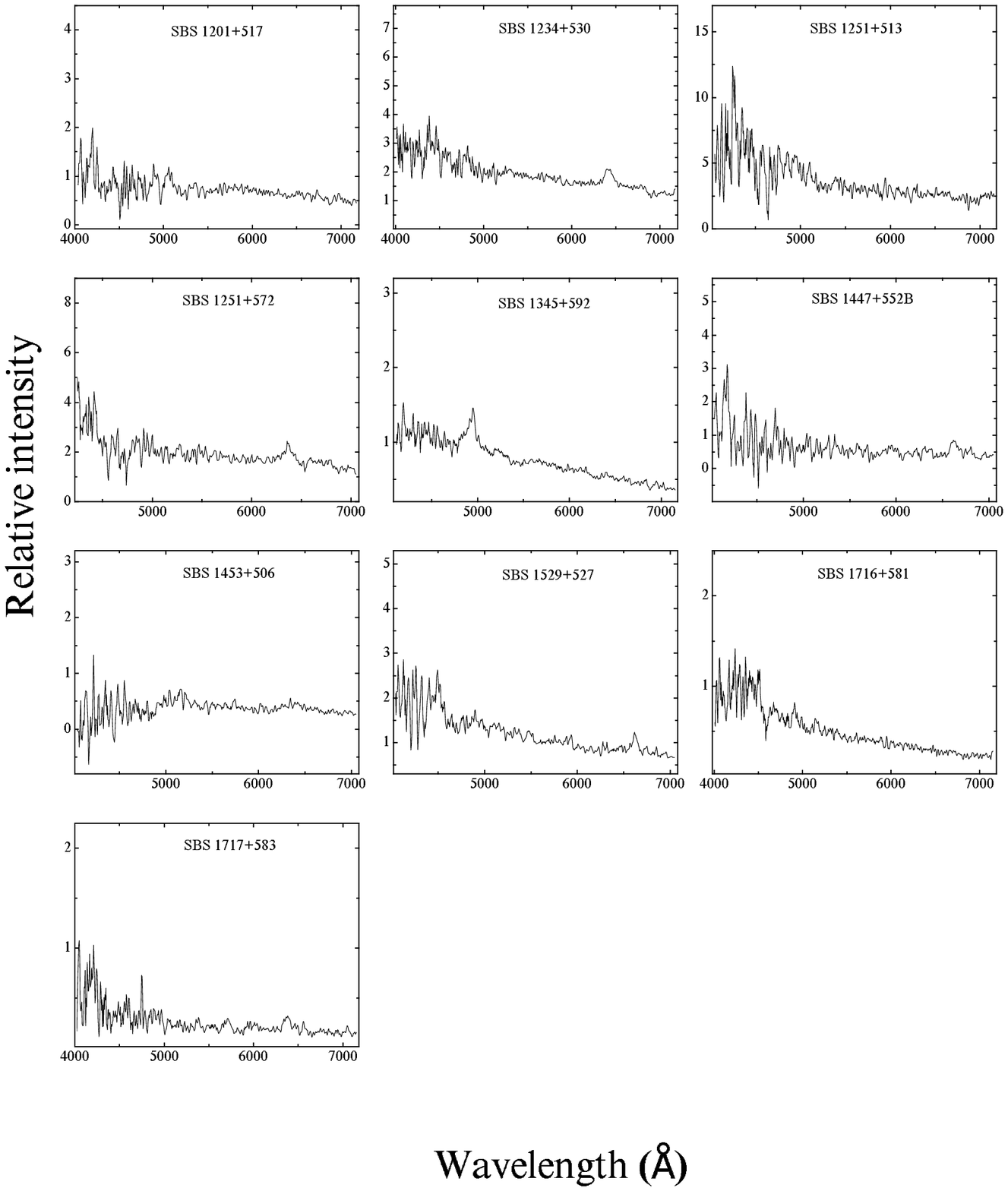}
\end{figure}

\begin{figure}[htb]
\setcounter{figure}{3}
\caption{Plots of the spectra of stars observed with the 2.1 m telescope.The vertical axis represents the relative flux the absissae represents the wavelength in \AA }

\epsfxsize=15cm\epsfbox{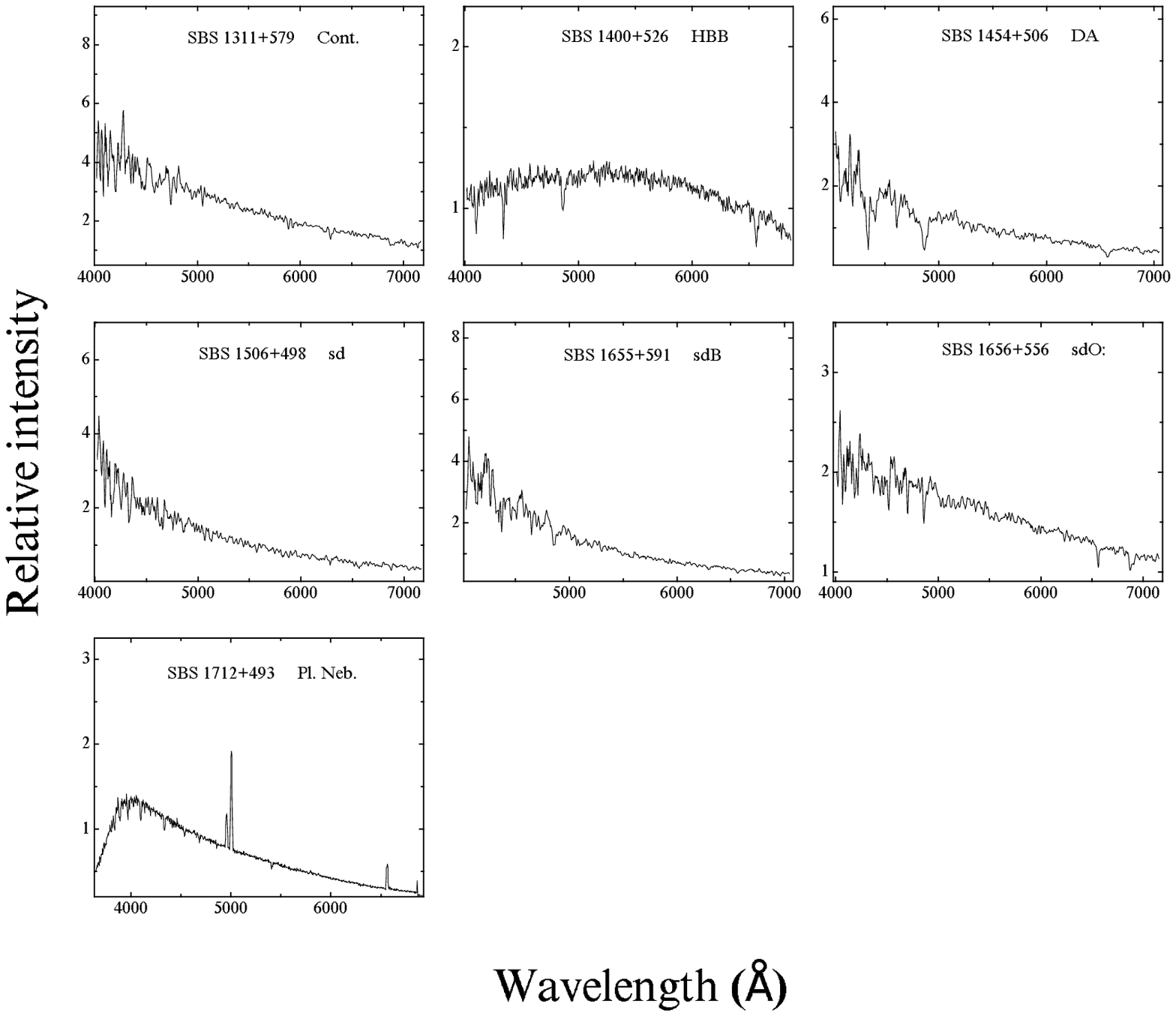}
\end{figure}

\clearpage

{\scriptsize
\tabcolsep=3mm
\begin{planotable}{lllllrrll}
\tablewidth{165mm}
\tablecaption{Journal of Observations}
\tablehead{
\colhead{SBS}            & \colhead{R.A.}   &
\colhead{Dec.}           & \colhead{m}      &
\colhead{Spect.  }       & \colhead{Date of} &
\colhead{Exp.}           & \colhead{Instru-} &
\colhead{Other}
\\[1ex]
\colhead{designation}    & \colhead{1950.0}   &
\colhead{1950.0}         & \colhead{pg}       &
\colhead{type}           & \colhead{obs.}     &
\colhead{sec}            & \colhead{ment}     &
\colhead{designation}}
\startdata
0749+590  & 07 49 10.34 & +59 02 27.5 & 16.0 & DA    & 07.02.83 &  420 &  I  &                \nl
0749+583  & 07 49 35.34 & +58 22 50.6 & 17.5 & G     & 02.11.91 &  600 &  II &       Mkn 381  \nl
0750+581  & 07 50 24.97 & +58 06 17.8 & 17.5 & DAF   & 14.01.93 &  879 &  II &                \nl
0753+610A & 07 53 08.89 & +61 02 17.6 & 17.5 & G     & 25 11 81 &  600 &  I  &                \nl
0755+600  & 07 55 11.93 & +60 02 02.6 & 17.5 & CV    & 25.11.81 &  960 &  I  &                \nl
0755+515  & 07 55 46.58 & +51 34 48.7 & 17.0 & NHB   & 25.02.92 &  407 &  II &                \nl
0756+508  & 07 56 18.91 & +50 50 40.7 & 17   & G     & 13.04.96 & 1200 &  IV &                \nl
0756+581  & 07 56 32.40 & +58 10 50.2 & 16.5 & sdB   & 14.01.93 &  583 &  II &                \nl
0756+566  & 07 56 32.74 & +56 41 51.2 & 17.0 & DA    & 16.11.90 &  414 &  II &                \nl
0759+608  & 07 59 20.79 & +60 53 54.1 & 18.0 & DA:   & 25.11.81 &  840 &  I  &                \nl
0759+610  & 07 59 21.51 & +61 02 12.0 & 18.0 & sdB   & 25.11.81 &  840 &  I  &                \nl
0800+491  & 08 00 41.15 & +49 09 08.6 & 11.0 & NHB   & 06.11.91 & 2297 &  II &                \nl
0801+537  & 08 01 29.66 & +53 43 33.9 & 17.0 & F     & 16.11.90 &  900 &  II &                \nl
0801+581  & 08 01 35.73 & +58 11 35.2 & 17   & QSO   & 14.01.93 &  745 &  II &                \nl
0803+510  & 08 03 16.13 & +51 03 57.9 & 16.0 & sdO   & 09.04.91 &  148 &  II &                \nl
0804+590  & 08 04 11.37 & +59 01 07.5 & 18.5 & sdB   & 25.11.81 &  960 &  I  &                \nl
0806+505  & 08 06 20.26 & +50 34 30.2 & 17.01& QSO   & 07.11.91 &  457 &  II &  RGBJ0810+504\tablenotemark{\S} \nl
0809+566  & 08 09 41.39 & +56 36 14.1 & 14.0 & DA    & 04.02.89 &  300 &  II &                \nl
0809+593  & 08 09 53.82 & +59 22 08.3 & 17.5 & DA    & 25.11.81 &  480 &  I  &                \nl
0810+524  & 08 10 19.80 & +52 26 22.1 & 17.5 & DA    & 09.04.91 &  236 &  II &                \nl
0811+582A & 08 11 10.23 & +58 12 43.3 & 18.0 & Cont  & 25.11.81 &  480 &  I  &                \nl
0811+513  & 08 11 31.10 & +51 22 25.6 & 17.5 & Cont  & 09.04.91 &  358 &  II &                \nl
0811+582B & 08 11 43.45 & +58 12 47.9 & 18   & NHB   & 08.02.97 & 1200 &  III&                \nl
0818+498  & 08 18 42.08 & +49 52 23.2 & 17.5 & DA:   & 08.11.91 &  252 &  II &                \nl
0821+602  & 08 21 00.35 & +60 13 44.1 & 16.5 & DA    & 09.04.91 &  265 &  II &                \nl
0822+552  & 08 22 41.31 & +55 15 45.6 & 16.5 & G     & 22.02.87 &  340 &  II &                \nl
0825+591  & 08 25 54.46 & +59 06 51.7 & 17.5 & DA    & 09.10.88 &  540 &  II &                \nl
0825+568  & 08 25 54.68 & +56 51 17.2 & 17.0 & sdO:  & 09.04.91 &  244 &  II &                \nl
0826+569  & 08 26 39.84 & +56 55 08.0 & 17.0 & DB    & 07.11.91 &  230 &  II &                \nl
0828+490  & 08 28 57.53 & +49 02 13.8 & 17.0 & sdB   & 08.11.91 &  162 &  II &                \nl
0829+559  & 08 29 42.74 & +55 57 46.9 & 16.5 & DB:   & 08.11.91 &  453 &  II &                \nl
0830+537  & 08 30 35.96 & +53 46 34.3 & 16.5 & DA    & 29.03.87 &  400 &  II &                \nl
0832+536  & 08 32 58.55 & +53 39 02.3 & 16.0 & CV    & 30.12.86 &  340 &  II &                \nl
0833+491  & 08 33 34.18 & +49 07 27.7 & 15.5 & DAF   & 06.11.91 & 2297 &  II &                \nl
0834+576  & 08 34 34.81 & +57 37 22.1 & 16.5 & DA+G  & 09.04.91 &  173 &  II &                \nl
0836+533  & 08 36 40.48 & +53 21 33.1 & 15.5 & DA    & 04.02.89 &  200 &  II &                \nl
0838+563  & 08 38 46.24 & +56 18 07.0 & 16.0 & sdB   & 04.02.89 &  340 &  II &                \nl
0839+541  & 08 39 29.41 & +54 08 07.2 & 16.96& QSO   & 08.04.91 &  377 &  II &                \nl
0841+565  & 08 41 55.42 & +56 33 14.0 & 17.0 & Cont  & 08.11.91 &  605 &  II &                \nl
0842+572  & 08 42 10.83 & +57 14 28.5 & 17.0 & DA    & 09.04.91 &  280 &  II &                \nl
0842+584  & 08 42 28.75 & +58 27 53.7 & 18   & QSO   & 17.02.94 & 1232 &  II &                \nl
0851+586  & 08 51 13.56 & +58 36 59.7 & 17.5 & G     & 10.11.91 &  587 &  II &                \nl
0853+506  & 08 53 02.09 & +50 39 21.2 & 17.0 & sd:   & 07.11.91 &  196 &  II &                \nl
0856+508  & 08 56 12.01 & +50 52 33.2 & 17.0 & sd    & 07.11.91 &  208 &  II &                \nl
0902+561  & 09 02 25.56 & +56 09 31.1 & 15.5 & DA    & 05.02.89 &  300 &  II &                \nl
0904+566  & 09 04 39.77 & +56 37 32.5 & 17.0 & G     & 07.11.91 &  215 &  II &                \nl
0905+549  & 09 05 51.52 & +54 58 07.6 & 16.5 & sdB   & 08.02.89 &  340 &  II &                \nl
0906+532  & 09 06 06.09 & +53 16 25.1 & 15.0 & sdB   & 25.02.88 &  300 &  II &                \nl
0906+552  & 09 06 57.66 & +55 17 41.3 & 15.5 & DA    & 05.02.89 &  300 &  II &                \nl
0910+584  & 09 10 47.66 & +58 25 04.9 & 17.0 & sdOB  & 07.11.91 &  195 &  II &                \nl
0911+527  & 09 11 36.43 & +52 44 22.9 & 17.0 & DA:   & 10.11.91 &  169 &  II &                \nl
0913+545  & 09 13 31.49 & +54 30 50.8 & 11.0 & HBB   & 25.05.88 &  120 &  II &                \nl
0914+546  & 09 14 59.12 & +54 40 59.1 & 13.0 & DAB   & 22.02.87 &  150 &  II &                \nl
0916+513  & 09 16 29.83 & +51 18 53.2 & 16.33& QSO   & 07.11.91 &  176 &  II &   NGC 2841 UB3 \nl
0919+529  & 09 19 22.29 & +52 57 43.8 & 16.0 & DA    & 05.03.92 &  192 &  II &                \nl
0920+597  & 09 20 18.66 & +59 44 22.0 & 17.5 & DA    & 15.11.79 &  420 &  I  &                \nl
0920+544  & 09 20 27.88 & +54 27 28.1 & 16.5 & F:    & 08.11.91 &  974 &  II &                \nl
0921+547  & 09 21 21.31 & +54 47 19.7 & 16.5 & DA    & 08.11.91 &  508 &  II &                \nl
0926+498  & 09 26 27.73 & +49 49 24.2 & 17.0 & DB    & 05.03.92 &  294 &  II &                \nl
0926+581  & 09 26 59.77 & +58 09 46.1 & 17.5 & G     & 08.01.92 &  434 &  II &                \nl
0927+540A & 09 27 04.69 & +54 04 09.1 & 17.5 & F     & 27.03.87 &  540 &  II &                \nl
0927+575  & 09 27 10.59 & +57 34 38.5 & 17.5 & NHB   & 08.01.92 &  707 &  II &                \nl
0927+540B & 09 27 26.10 & +54 03 47.5 & 16.5 & F     & 27.03.87 &  500 &  II &                \nl
0928+559  & 09 28 26.21 & +55 55 52.9 & 16.0 & G     & 22.02.87 &  480 &  II &                \nl
0929+593  & 09 29 12.79 & +59 19 16.1 & 17.5 & G     & 08.01.92 &  344 &  II &                \nl
0929+556  & 09 29 34.12 & +55 39 41.6 & 16.5 & G:    & 08.11.91 &  679 &  II &                \nl
0933+579  & 09 33 24.87 & +57 58 55.1 & 17.5 & G     & 08.01.92 & 1219 &  II &                \nl
0933+515  & 09 33 59.58 & +51 31 27.7 & 17.5 & G:    & 08.01.92 &  423 &  II &                \nl
0934+590  & 09 34 12.80 & +59 04 43.5 & 17.5 & G     & 08.01.92 &  300 &  II &                \nl
0934+499  & 09 34 34.64 & +49 55 22.5 & 17.0 & DA    & 10.02.86 &  320 &  II &                \nl
0936+495  & 09 36 52.10 & +49 34 39.3 & 18.5 & DA    & 10.02.86 &  900 &  II &                \nl
0937+583  & 09 37 05.67 & +58 23 27.6 & 17.5 & G     & 08.01.92 &  329 &  II &                \nl
0937+510  & 09 37 06.87 & +51 01 29.8 & 18.0 & sd    & 10.11.85 &  900 &  II &                \nl
0937+503  & 09 37 31.42 & +50 22 57.0 & 18.5 & QSO   & 06.04.86 &  960 &  II &                \nl
0937+552  & 09 37 48.98 & +55 13 46.6 & 18.5 & F     & 17.03.80 &  960 &  I  &                \nl
0937+519  & 09 37 57.95 & +51 56 42.5 & 17.5 & DG    & 10.02.86 &  480 &  II &                \nl
0938+533  & 09 38 06.37 & +53 22 25.3 & 18.0 & G     & 17.03.80 &  600 &  I  &                \nl
0938+550A & 09 38 35.01 & +55 00 13.0 & 18.0 & DA    & 17.03.80 &  660 &  I  &                \nl
0938+605  & 09 38 35.22 & +60 30 15.8 & 17.5 & G     & 06.04.91 &  197 &  II &                \nl
0938+577  & 09 38 38.83 & +57 47 24.1 & 17.5 & DA    & 14.11.79 &  380 &  I  &                \nl
0938+573  & 09 38 56.66 & +57 18 45.9 & 17.0 & G     & 10.11.91 &  513 &  II &                \nl
0939+548  & 09 39 14.24 & +54 49 06.9 & 18.0 & G     & 17.03.80 &  780 &  I  &                \nl
0940+534  & 09 40 16.08 & +53 29 21.0 & 18.5 & DA    & 17.03.80 &  900 &  I  &                \nl
0940+512A & 09 40 51.68 & +51 12 32.7 & 18.0 & G     & 05.04.86 &  840 &  II &                \nl
0941+537  & 09 41 17.52 & +53 42 40.5 & 17.5 & G     & 17.03.80 &  540 &  I  &                \nl
0941+551  & 09 41 31.36 & +55 08 40.3 & 17.5 & DA    & 03.01.78 &  720 &  I  &                \nl
0941+558  & 09 41 41.31 & +55 48 39.9 & 12.0 & DAF:  & 05.02.89 &  120 &  II &                \nl
0941+514  & 09 41 55.73 & +51 27 23.9 & 18.5 & DAO   & 14.02.86 &  900 &  II &                \nl
0942+507  & 09 42 18.55 & +50 44 28.5 & 19.0 & DAO   & 13.02.86 &  960 &  II &                \nl
0942+527B & 09 42 51.17 & +52 47 34.0 & 18.5 & DA:   & 29.11.87 &  900 &  II &                \nl
0943+507A & 09 43 11.37 & +50 42 07.1 & 14.5 & G     & 15.12.87 &  300 &  II &                \nl
0943+603  & 09 43 17.24 & +60 20 16.7 & 17.5 & DA    & 15.11.79 &  600 &  I  &                \nl
0943+592  & 09 43 48.45 & +59 13 18.3 & 17.0 & G     & 10.11.91 &  257 &  II &                \nl
0944+506  & 09 44 14.28 & +50 38 40.0 & 18.0 & DA    & 10.02.86 &  780 &  II &                \nl
0944+560  & 09 44 59.91 & +56 01 06.7 & 18.0 & DAF:  & 17.03.80 &  780 &  I  &                \nl
0945+578  & 09 45 21.99 & +57 53 57.4 & 17.5 & DA:   & 08.01.92 &  665 &  II &                \nl
0946+501B & 09 46 48.11 & +50 09 50.6 & 17.0 & DA    & 10.02.86 &  640 &  II &                \nl
0947+549  & 09 47 15.97 & +54 55 33.0 & 18.0 & sd:   & 17.03.80 &  800 &  I  &                \nl
0948+513  & 09 48 10.41 & +51 23 52.0 & 18.5 & sd    & 26.11.87 &  860 &  II &                \nl
0948+550  & 09 48 17.30 & +55 04 26.7 & 17.0 & F     & 10.11.91 &  421 &  II &                \nl
0948+505  & 09 48 24.95 & +50 31 21.1 & 18.0 & DBA   & 10.02.86 &  800 &  II &                \nl
0949+554  & 09 49 39.49 & +55 25 54.1 & 16.0 & F     & 19.03.80 &  540 &  I  &                \nl
0950+579  & 09 50 07.23 & +57 56 07.5 & 16.5 & sdB   & 10.11.91 &  227 &  II &                \nl
0950+562  & 09 50 34.51 & +56 12 02.7 & 19.0 & DA    & 28.12.84 &  900 &  II &                \nl
0950+568  & 09 50 39.47 & +56 48 48.9 & 17.5 & G     & 08.02.83 &  780 &  I  &                \nl
0950+575  & 09 50 47.40 & +57 33 51.4 & 17.0 & G     & 08.01.92 &  292 &  II &                \nl
0951+591  & 09 51 26.40 & +59 07 46.3 & 16.5 & G     & 08.01.92 &  325 &  II &                \nl
0951+497  & 09 51 45.83 & +49 42 50.6 & 18.0 & G     & 10.10.88 &  860 &  II &                \nl
0953+574  & 09 53 14.84 & +57 27 41.4 & 17.5 & DA    & 06.04.91 &  294 &  II &                \nl
0953+509  & 09 53 16.87 & +50 56 41.3 & 18.0 & G     & 09.03.88 &  860 &  II &                \nl
0955+606  & 09 55 13.14 & +60 36 08.8 & 16.5 & DAF   & 08.01.92 & 1141 &  II &                \nl
0955+524  & 09 55 27.99 & +52 29 19.7 & 18.0 & Cont  & 27.11.87 &  800 &  II &                \nl
0956+540  & 09 56 07.53 & +54 00 53.4 & 17.0 & G:    & 16.03.80 &  600 &  I  &                \nl
0956+492  & 09 56 10.26 & +49 12 47.7 & 17.5 & sdB   & 27.11.87 &  320 &  II &                \nl
0957+513  & 09 57 03.20 & +51 18 29.5 & 17.5 & sdB   & 10.02.86 &  360 &  II &                \nl
0957+553  & 09 57 21.35 & +55 21 21.9 & 17.0 & F     & 16.03.80 &  600 &  I  &                \nl
0957+551  & 09 57 57.62 & +55 06 05.3 & 17.5 & DAF:  & 16.03.80 &  720 &  I  &                \nl
0958+532  & 09 58 06.50 & +53 15 41.7 & 18.0 & DAF:  & 16.03.80 &  780 &  I  &                \nl
0958+610  & 09 58 30.43 & +61 03 36.8 & 16.5 & DA    & 29.02.92 &  274 &  II &                \nl
0958+580  & 09 58 56.06 & +58 04 40.4 & 17.5 & G     & 06.04.91 &  252 &  II &                \nl
1001+559  & 10 01 27.00 & +55 58 33.1 & 17.0 & G     & 16.03.80 &  780 &  I  &                \nl
1002+562  & 10 02 47.56 & +56 15 21.4 & 17.0 & DAO   & 22.02.90 &  360 &  II &                \nl
1003+606  & 10 03 51.36 & +60 41 38.7 & 16.5 & G     & 09.01.92 &  237 &  II &                \nl
1004+598  & 10 04 21.39 & +59 52 30.3 & 16.5 & G     & 14.11.79 &  360 &  I  &                \nl
1004+572  & 10 04 21.81 & +57 17 03.5 & 17.5 & F     & 09.04.91 &  523 &  II &                \nl
1005+584  & 10 05 04.56 & +58 24 35.7 & 17.5 & G     & 09.04.91 &  188 &  II &                \nl
1006+599A & 10 06 51.00 & +59 54 40.0 & 17.0 & G     & 09.04.91 &  171 &  II &                \nl
1006+578B & 10 06 58.66 & +57 52 35.5 & 16.5 & G     & 14.11.79 &  600 &  I  &                \nl
1009+538  & 10 09 11.55 & +53 48 51.4 & 17.5 & DA    & 04.02.92 &  540 &  II &                \nl
1009+585  & 10 09 58.60 & +58 34 34.8 & 16.5 & sdB   & 09.04.91 &  194 &  II &                \nl
1013+565  & 10 13 18.74 & +56 30 18.6 & 18.0 & Pec * & 10.11.91 &  472 &  II &                \nl
1015+532  & 10 15 20.14 & +53 12 11.1 & 16.0 & F     & 05.03.92 &  468 &  II &                \nl
1016+510  & 10 16 07.16 & +51 01 01.4 & 17   & QSO   & 03.12.95 & 1200 & III &                \nl
1016+563A & 10 16 09.23 & +56 18 34.9 & 17.5 & DA:   & 04.02.92 &  560 &  II &                \nl
1016+527  & 10 16 11.90 & +52 47 07.5 & 16.5 & DA    & 05.03.92 &  973 &  II &                \nl
1017+533  & 10 17 14.66 & +53 19 39.7 & 17.0 & CV    & 26.02.88 &  540 &  II &                \nl
1018+601  & 10 18 56.21 & +60 07 12.9 & 17.5 & G     & 09.04.91 &  362 &  II &                \nl
1020+562  & 10 20 22.33 & +56 12 06.6 & 18.0 & G     & 07.11.91 &  364 &  II &                \nl
1020+553A & 10 20 22.81 & +55 21 15.7 & 16.5 & G     & 10.11.91 &  248 &  II &                \nl
1021+562  & 10 21 38.30 & +56 13 47.6 & 18.02& NHB   & 07.11.91 &  191 &  II &                \nl
1022+594  & 10 22 40.84 & +59 29 40.9 & 17.5 & DB    & 14.11.79 &  660 &  I  &                \nl
1025+576  & 10 25 59.96 & +57 39 15.5 & 17   & QSO   & 17.02.94 & 1800 &  II &                \nl
1026+560  & 10 26 48.89 & +56 02 29.7 & 18.0 & DA    & 07.03.88 &  720 &  II &                \nl
1027+500  & 10 27 12.45 & +50 00 44.3 & 16.5 & sd    & 10.11.91 &  212 &  II &                \nl
1029+537  & 10 29 02.04 & +53 45 03.3 & 13.5 & DA    & 23.03.87 &  180 &  II &   \nl
1029+526  & 10 29 08.84 & +52 36 49.0 & 16   & sd    & 15.04.96 & 1200 &  IV &                \nl
1033+571  & 10 33 29.72 & +57 07 00.6 & 17.39& QSO   & 10.11.91 &  210 &  II &                \nl
1034+496  & 10 34 23.80 & +49 41 17.2 & 16.5 & sd:   & 10.11.91 &  239 &  II &                \nl
1034+557  & 10 34 56.52 & +55 46 58.0 & 17.0 & DA    & 09.01.92 &  231 &  II &                \nl
1035+541  & 10 35 57.60 & +54 10 28.5 & 17.5 & DA    & 09.01.92 &  251 &  II &                \nl
1036+550  & 10 36 22.15 & +55 05 53.3 & 17.5 & DA    & 09.01.92 &  252 &  II &                \nl
1040+493  & 10 40 11.87 & +49 18 07.1 & 16.5 & sd    & 29.02.92 &  277 &  II &   \nl
1040+520  & 10 40 48.77 & +52 00 54.1 & 17.5 & DA    & 09.01.92 &  217 &  II &                \nl
1043+569  & 10 43 48.09 & +56 59 47.2 & 17.5 & F     & 04.02.92 &  660 &  II &                \nl
1045+570  & 10 45 15.66 & +57 03 29.4 & 17.5 & DA    & 09.01.92 &  507 &  II &                \nl
1047+557A & 10 47 26.47 & +55 42 25.6 & 17.0 & DA    & 03.03.87 &  540 &  II &                \nl
1047+557B & 10 47 51.85 & +55 43 18.3 & 17.17& QSO   & 17.02.94 &  711 &  II &                \nl
1049+541  & 10 49 10.46 & +54 07 31.0 & 16.0 & DA    & 22.04.87 &  360 &  II &                \nl
1050+582  & 10 50 51.49 & +58 16 26.1 & 17.0 & DA    & 06.04.91 &  220 &  II &                \nl
1051+556  & 10 51 48.63 & +55 39 08.8 & 16.5 & DA:   & 10.11.91 &  114 &  II &                \nl
1053+561  & 10 53 26.66 & +56 11 16.1 & 17.5 & sd    & 09.01.92 &  343 &  II &                \nl
1056+516A & 10 56 20.09 & +51 40 48.9 & 15.5 & Cont  & 10.11.91 &  286 &  II &   \nl
1057+556  & 10 57 31.09 & +55 38 47.0 & 17.5 & Cont  & 08.01.91 &  540 &  II &                \nl
1058+570  & 10 58 10.81 & +57 05 31.5 & 17.5 & DA    & 25.02.88 &  660 &  II &                \nl
1058+561  & 10 58 38.79 & +56 08 18.2 & 18.84& QSO   & 17.02.93 & 1272 &  II &                \nl
1058+559  & 10 58 51.56 & +55 54 04.6 & 16.0 & DA    & 27.12.89 &  500 &  II &                \nl
1059+568  & 10 59 23.36 & +56 51 26.0 & 16.5 & DA:   & 09.01.92 &  324 &  II &                \nl
1101+525  & 11 01 18.91 & +52 30 46.3 & 17.5 & DA    & 08.01.92 &  508 &  II &                \nl
1102+558  & 11 02 51.61 & +55 52 19.8 & 17.0 & F     & 08.01.92 &  423 &  II &                \nl
1103+595  & 11 03 00.40 & +59 33 49.2 & 17.5 & Cont  & 13.03.85 &  660 &  II &                \nl
1103+511  & 11 03 38.35 & +51 09 17.6 & 17.5 & DAB:  & 09.01.92 &  176 &  II &                \nl
1103+586  & 11 03 38.48 & +58 41 08.2 & 17.5 & DA    & 09.01.92 &  269 &  II &                \nl
1107+603  & 11 07 11.20 & +60 18 00.1 & 18.0 & Cont  & 13.03.85 &  780 &  II &                \nl
1108+506  & 11 08 06.81 & +50 37 14.0 & 17.0 & sdB   & 09.01.92 &  620 &  II &                \nl
1108+540  & 11 08 46.12 & +54 04 10.3 & 17.0 & DA    & 03.01.78 &  660 &  I  &                \nl
1112+572  & 11 12 34.25 & +57 17 46.6 & 17.0 & DA    & 13.01.78 &  600 &  I  &                \nl
1113+554B & 11 13 54.05 & +55 27 50.1 & 17.0 & DA    & 04.02.92 &  540 &  II &                \nl
1116+518  & 11 16 49.04 & +51 49 41.9 & 16.98& QSO   & 26.04.87 &  600 &  II &                \nl
1124+612  & 11 24 05.11 & +61 17 21.1 & 17.5 & DA    & 08.02.83 &  660 &  II &                \nl
1125+558  & 11 25 05.05 & +55 51 53.6 & 16.5 & DB    & 29.02.92 &  590 &  II &                \nl
1125+596  & 11 25 15.52 & +59 36 28.2 & 16.5 & DA    & 13.01.78 &  540 &  I  &                \nl
1127+512  & 11 27 09.68 & +51 16 33.1 & 17.0 & sdO:  & 05.03.92 &  597 &  II &                \nl
1128+499  & 11 28 16.17 & +49 54 59.9 & 16.0 & DA    & 29.02.92 &  242 &  II &                \nl
1131+521  & 11 31 13.87 & +52 08 40.3 & 17.0 & DA:   & 05.03.92 &  325 &  II &                \nl
1133+489  & 11 33 27.49 & +48 59 55.7 & 16.5 & sdOB  & 29.02.92 &  252 &  II &                \nl
1139+583  & 11 39 53.94 & +58 22 28.0 & 18.5 & G     & 25.11.81 &  780 &  I  &                \nl
1142+570  & 11 42 53.63 & +57 00 26.4 & 14.5 & HBB   & 22.04.87 &  220 &  II &                \nl
1144+603  & 11 44 12.11 & +60 20 16.2 & 18.0 & F     & 25.11.81 &  720 &  I  &                \nl
1144+599  & 11 44 51.61 & +59 55 59.9 & 17.0 & DA    & 13.01.78 &  660 &  I  &                \nl
1148+564  & 11 48 00.01 & +56 25 27.1 & 15.0 & DA    & 28.03.87 &  300 &  II &                \nl
1149+598  & 11 49 03.63 & +59 51 14.7 & 18.5 & Cont  & 29.03.86 &  900 &  II &                \nl
1149+560  & 11 49 33.46 & +56 04 48.1 & 16.0 & DAF:  & 28.03.87 &  400 &  II &                \nl
1149+509  & 11 49 57.05 & +50 56 37.8 & 15.0 & DA:   & 09.04.91 &  280 &  II &                \nl
1150+599A & 11 50 47.02 & +59 56 38.5 & 17.5 & CV    & 04.04.86 &  660 &  II &                \nl
1151+587  & 11 51 58.87 & +58 46 37.9 & 17.0 & DBA   & 13.01.78 &  540 &  I  &                \nl
1154+555  & 11 54 19.52 & +55 34 20.7 & 16.0 & HBB   & 29.03.87 &  420 &  II &                \nl
1154+583A & 11 54 30.32 & +58 21 21.1 & 18.0 & DA    & 04.04.86 &  720 &  II &                \nl
1154+514  & 11 54 47.37 & +51 26 54.5 & 16.5 & DA    & 25.03.92 &  600 &  II &                \nl
1154+561  & 11 54 53.74 & +56 11 50.6 & 16.5 & G     & 29.03.89 &  600 &  II &                \nl
1155+594  & 11 55 52.75 & +59 26 08.4 & 17.0 & DA    & 13.01.78 &  600 &  I  &                \nl
1158+597  & 11 58 10.20 & +59 42 40.3 & 17.5 & DB    & 13.01.78 &  420 &  I  &                \nl
1158+538  & 11 58 36.65 & +53 53 46.4 & 18.66& QSO   & 13.01.78 &  900 &  I  &                \nl
1158+599  & 11 58 58.92 & +59 57 17.7 & 17.5 & DA    & 13.12.85 &  600 &  II &                \nl
1200+589A & 12 00 40.49 & +58 57 12.8 & 17.0 & F     & 04.04.86 &  540 &  II &                \nl
1201+517  & 12 01 00.31 & +51 46 56.8 & 17.24& QSO   & 06.03.97 & 2400 &  IV &                \nl
1203+587  & 12 03 19.86 & +58 46 40.9 & 18.5 & DA    & 04.04.86 &  780 &  II &                \nl
1204+560  & 12 04 23.47 & +56 03 29.6 & 17.0 & DA    & 06.04.91 &  295 &  II &                \nl
1210+511  & 12 10 00.26 & +51 10 43.2 & 17.0 & sd    & 25.02.92 &  742 &  II &                \nl
1210+537B & 12 10 28.56 & +53 43 37.4 & 18.0 & sdB-O & 16.03.80 &  720 &  I  &                \nl
1215+552  & 12 15 49.03 & +55 14 47.0 & 19.5 & Cont  & 27.12.84 &  900 &  II &                \nl
1216+522  & 12 16 38.32 & +52 12 10.9 & 17.5 & F     & 19.05.93 &  660 &  II &                \nl
1216+610  & 12 16 57.86 & +61 01 19.8 & 17.5 & G     & 11.04.91 &  258 &  II &                \nl
1217+490  & 12 17 06.98 & +49 02 13.1 & 17   & NHB   & 12.03.97 & 3600 &  IV &                \nl
1217+559B & 12 17 55.22 & +55 59 47.2 & 18.0 & F     & 19.02.82 &  720 &  I  &                \nl
1218+538  & 12 18 56.34 & +53 53 59.9 & 19.5 & sd    & 20.02.82 &  900 &  I  &                \nl
1221+537  & 12 21 12.02 & +53 45 07.1 & 19.0 & DA    & 17.03.86 &  960 &  II &                \nl
1223+533A & 12 23 14.95 & +53 18 52.6 & 18.5 & F:    & 22.02.82 &  900 &  I  &                \nl
1224+569  & 12 24 55.39 & +56 55 02.0 & 19.0 & DA    & 20.02.82 &  900 &  I  &                \nl
1226+570  & 12 26 22.38 & +57 01 36.1 & 18.5 & sdB   & 20.02.82 &  780 &  I  &                \nl
1227+553  & 12 27 15.19 & +55 22 49.7 & 16.5 & Cont  & 16.03.80 &  540 &  I  &                \nl
1228+551  & 12 28 15.22 & +55 07 33.1 & 18.0 & F:    & 19.02.82 &  780 &  I  &                \nl
1229+580  & 12 29 52.54 & +58 03 41.7 & 17.0 & DA    & 11.04.91 &  140 &  II &                \nl
1231+494  & 12 31 13.05 & +49 27 00.5 & 16.5 & sdB   & 16.04.96 &  900 &  IV &                \nl
1233+523  & 12 33 20.92 & +52 22 42.5 & 17.0 & DA    & 25.03.92 &  660 &  II &                \nl
1234+530  & 12 34 19.72 & +53 05 13.4 & 16.5 & QSO   & 12.03.97 & 2400 &  IV &                \nl
1239+508  & 12 39 37.75 & +50 52 27.8 & 16.5 & DA    & 05.03.92 &  566 &  II &                \nl
1240+507  & 12 40 45.24 & +50 43 52.6 & 17.0 & sdB   & 02.04.92 &  660 &  II &                \nl
1241+562  & 12 41 24.64 & +56 12 30.5 & 17.5 & G     & 07.03.88 &  720 &  II &                \nl
1241+586  & 12 41 54.34 & +58 40 17.0 & 17.5 & DAO   & 11.04.91 &  280 &  II &                \nl
1244+566  & 12 44 31.49 & +56 36 24.5 & 17.0 & DA    & 11.04.91 &  160 &  II &                \nl
1245+567  & 12 45 14.78 & +56 46 21.7 & 16.5 & sd    & 11.01.91 &  207 &  II &                \nl
1247+568  & 12 47 04.63 & +56 48 35.9 & 17.5 & DA    & 11.04.91 &  230 &  II &                \nl
1251+572  & 12 51 14.65 & +57 17 32.8 & 17.83& QSO   & 11.04.97 & 3600 &  IV &                \nl
1251+513  & 12 51 24.76 & +51 18 20.1 & 16   & QSO   & 16.04.96 & 1200 &  IV &                \nl
1251+586  & 12 51 51.32 & +58 36 03.0 & 18.0 & DA    & 14.05.85 &  780 &  II &                \nl
1257+609  & 12 57 34.73 & +60 55 08.7 & 16.5 & sdB   & 05.03.92 &  233 &  II &                \nl
1258+585  & 12 58 11.30 & +58 32 04.6 & 18.0 & QSO   & 18.05.93 & 1314 &  II &                \nl
1300+523  & 13 00 24.33 & +52 23 19.4 & 16.0 & DAB   & 05.03.92 &  144 &  II &                \nl
1300+514  & 13 00 25.19 & +51 26 01.5 & 16.5 & G     & 13.03.97 & 2400 &  IV &                \nl
1303+536  & 13 03 30.90 & +53 38 11.2 & 17.0 & DA    & 06.04.91 &  150 &  II &                \nl
1303+565  & 13 03 57.21 & +56 31 39.2 & 17.0 & DA:   & 26.04.87 &  780 &  II &                \nl
1304+541  & 13 04 27.57 & +54 06 09.8 & 17.0 & DA    & 02.04.91 &  780 &  II &                \nl
1305+538  & 13 05 21.41 & +53 51 53.3 & 17.5 & QSO   & 18.05.93 &  809 &  II &                \nl
1306+563  & 13 06 25.52 & +56 21 36.5 & 17.0 & sdB   & 02.04.91 &  660 &  II &                \nl
1307+562  & 13 07 05.18 & +56 13 35.5 & 18   & QSO   & 17.05.93 & 1215 &  II &                \nl
1309+544  & 13 09 42.68 & +54 27 00.6 & 17.0 & DA+K  & 10.04.91 &  279 &  II &                \nl
1309+511  & 13 09 58.37 & +51 09 23.5 & 16.0 & DA:   & 06.03.92 &  469 &  II &                \nl
1311+504  & 13 11 31.65 & +50 24 27.2 & 16.0 & HBB   & 11.04.96 &  900 &  IV &                \nl
1311+579  & 13 11 37.45 & +57 54 09.8 & 16   & Cont  & 06.03.97 & 2400 &  IV &                \nl
1314+537  & 13 14 36.41 & +53 43 03.9 & 16.5 & DA:   & 06.03.89 &  540 &  II &                \nl
1315+605  & 13 15 20.15 & +60 31 20.2 & 18   & QSO   & 17.02.94 & 1800 &  III&                \nl
1317+526  & 13 17 24.19 & +52 39 19.3 & 17.0 & DA    & 04.04.92 &  660 &  II &                \nl
1319+555  & 13 19 17.59 & +55 32 25.1 & 17.0 & DA    & 26.02.88 &  660 &  II &                \nl
1321+496  & 13 21 54.85 & +49 38 07.7 & 12.0 & sdB   & 20.05.93 &  310 &  II &                \nl
1330+580  & 13 30 06.44 & +58 00 34.2 & 16.5 & DB    & 20.03.91 &  155 &  II &                \nl
1337+570  & 13 37 55.75 & +57 00 15.0 & 17.5 & DA    & 20.03.91 &  410 &  II &                \nl
1338+551  & 13 38 48.77 & +55 11 32.6 & 17.5 & QSO   & 17.05.93 & 1280 &  II &                \nl
1339+606  & 13 39 16.28 & +60 41 19.7 & 17.0 & DA    & 24.03.92 &  660 &  II &                \nl
1340+575  & 13 40 39.15 & +57 35 23.7 & 17.5 & DA:   & 15.04.91 &  999 &  II &                \nl
1345+592  & 13 45 35.77 & +59 17 28.6 & 16.5 & QSO   & 16.04.96 & 1200 &  IV &                \nl
1347+539A & 13 47 31.11 & +53 58 13.4 & 16   & NHB   & 10.04.96 &  900 &  IV &                \nl
1352+542  & 13 52 44.70 & +54 15 58.9 & 17.5 & F     & 10.04.91 &  320 &  II &                \nl
1353+538  & 13 53 25.69 & +53 49 22.1 & 11.0 & sdOA  & 06.03.89 &  120 &  II &                \nl
1356+564  & 13 56 06.80 & +56 25 36.5 & 17.0 & DA    & 06.03.92 &  235 &  II &                \nl
1359+506  & 13 59 07.36 & +50 39 56.4 & 17.0 & HBB   & 15.04.91 &  416 &  II &                \nl
1359+521B & 13 59 22.28 & +52 09 59.3 & 16.5 & NHB   & 06.03.92 &  186 &  II &                \nl
1400+526  & 14 00 40.59 & +52 36 41.5 & 15.5 & HBB   & 11.04.96 &  480 &  IV &                \nl
1402+529  & 14 02 50.39 & +52 57 47.7 & 17.0 & F:    & 11.07.91 &  829 &  II &                \nl
1403+535  & 14 03 55.48 & +53 30 09.1 & 17   & sd    & 14.04.96 & 1800 &  IV &                \nl
1407+521  & 14 07 46.44 & +52 08 07.6 & 17.0 & DA    & 24.03.92 &  780 &  II &                \nl
1411+546B & 14 11 35.24 & +54 37 40.5 & 17.0 & Cont  & 20.03.91 &  630 &  II & CBS 259\tablenotemark{\dag} \nl
1412+542  & 14 12 37.29 & +54 17 45.5 & 16.5 & DA    & 28.02.92 &  167 &  II & CBS 260\tablenotemark{\dag} \nl
1415+499  & 14 15 35.36 & +49 55 17.9 & 17.5 & NHB   & 12.04.96 & 1800 &  IV & CSO 627\tablenotemark{\dag} \nl
1415+573  & 14 15 35.75 & +57 21 06.3 & 16.5 & NHB   & 28.02.92 &  160 &  II &                             \nl
1416+519  & 14 16 21.20 & +51 57 43.6 & 18.0 & DA    & 28.02.92 &  355 &  II &                             \nl
1417+510  & 14 17 24.32 & +51 00 44.6 & 17.5 & QSO   & 11.02.97 &  600 & III &                             \nl
1417+514  & 14 17 51.53 & +51 28 28.1 & 18   & QSO   & 04.06.94 &  902 &  II & CSO 632\tablenotemark{\dag} \nl
1418+524  & 14 18 36.62 & +52 29 30.7 & 17.0 & DA:   & 28.02.92 &  157 &  II & CBS 264\tablenotemark{\dag} \nl
1422+589B & 14 22 27.88 & +58 58 49.2 & 17.0 & G     & 28.02.92 &  222 &  II &                             \nl
1422+497  & 14 22 53.15 & +49 43 29.5 & 16.0 & DA    & 06.04.92 &  470 &  II & CSO 645\tablenotemark{\dag} \nl
1423+500  & 14 23 13.15 & +50 00 59.1 & 18   & QSO   & 04.06.94 &  611 &  II & CSO 646\tablenotemark{\dag} \nl
1424+502B & 14 24 41.74 & +50 16 13.5 & 18   & QSO   & 17.02.94 &  683 &  II & CSO 647\tablenotemark{\dag} \nl
1425+578  & 14 25 59.61 & +57 52 34.9 & 17.5 & HBB   & 19.05.93 &  240 &  II &                             \nl
1426+574  & 14 26 53.47 & +57 24 20.0 & 17.5 & F     & 15.04.91 &  485 &  II &                             \nl
1428+490A & 14 28 07.49 & +49 01 53.6 & 17.5 & sdB   & 20.05.93 &  170 &  II & CBS 270\tablenotemark{\dag} \nl
1428+567  & 14 28 24.47 & +56 45 03.6 & 16.0 & sdB   & 11.04.91 & 1022 &  II &                             \nl
1428+490B & 14 28 34.16 & +49 04 55.2 & 14.0 & sdB   & 20.05.93 &  122 &  II & CBS 272\tablenotemark{\dag} \nl
1429+513  & 14 29 04.32 & +51 20 43.9 & 17.5 & DA    & 13.04.96 & 2400 &  IV & CSO 662\tablenotemark{\dag} \nl
1430+499  & 14 30 46.03 & +49 57 20.1 & 17.5 & NHB:  & 12.03.97 & 3600 &  IV &                             \nl
1432+507  & 14 32 46.66 & +50 42 28.0 & 16.5 & NHB   & 14.04.96 & 1800 &  IV &                             \nl
1432+504  & 14 32 47.28 & +50 24 07.5 & 17.0 & F     & 05.04.92 &  780 &  II &                             \nl
1433+510  & 14 33 32.49 & +51 05 25.8 & 17.0 & DA    & 05.04.92 &  720 &  II &                             \nl
1434+592B & 14 34 56.43 & +59 16 57.1 & 17.5 & F     & 14.05.85 &  780 &  II &                             \nl
1435+500A & 14 35 04.03 & +50 05 54.5 & 17.5 & QSO   & 04.06.94 &  760 &  II & CSO 673\tablenotemark{\dag} \nl
1437+526  & 14 37 54.89 & +52 47 39.7 & 17.0 & Pec.* & 06.03.97 & 1200 &  IV &                             \nl
1438+602B & 14 38 19.11 & +60 15 38.8 & 17.0 & NHB   & 05.04.92 &  720 &  II &                             \nl
1440+562  & 14 40 49.19 & +56 15 12.4 & 17.5 & Cont  & 20.03.91 &  328 &  II &                             \nl
1441+514  & 14 41 13.82 & +51 29 57.4 & 17.0 & NHB   & 05.04.92 &  720 &  II & CBS 280\tablenotemark{\dag} \nl
1442+495  & 14 42 03.82 & +49 30 13.4 & 17.0 & sdB   & 05.04.92 &  720 &  II & CBS 282\tablenotemark{\dag} \nl
1447+552B & 14 47 42.43 & +55 16 53.0 & 18.04& QSO   & 11.04.97 & 3600 &  IV &                             \nl
1449+537  & 14 49 34.11 & +53 46 11.6 & 19.35& QSO   & 11.02.97 & 1200 & III & RGBJ1451+535\tablenotemark{\S} \nl
1449+513  & 14 49 41.34 & +51 23 05.2 & 16.0 & DA    & 06.03.92 &  150 &  II & CBS 289\tablenotemark{\dag} \nl
1451+494  & 14 51 57.86 & +49 26 01.2 & 17.5 & NHB:  & 11.04.97 & 3600 &  IV &                             \nl
1451+606  & 14 51 58.13 & +60 36 20.4 & 17.5 & G     & 20.03.91 &  320 &  II &                             \nl
1452+553  & 14 52 48.57 & +55 23 58.4 & 16.0 & DA    & 28.01.90 &  600 &  II & CBS 295\tablenotemark{\dag} \nl
1453+506  & 14 53 10.14 & +50 41 16.6 & 19.04& QSO   & 12.04.97 & 3600 &  IV &                             \nl
1454+506  & 14 54 10.91 & +50 39 29.2 & 17.84& DA    & 14.04.97 & 3600 &  IV & CSO 700\tablenotemark{\dag} \nl
1500+520  & 15 00 34.06 & +52 03 50.1 & 17.0 & DA    & 10.07.91 &  560 &  II & CSO 715\tablenotemark{\dag} \nl
1500+531  & 15 00 49.35 & +53 10 37.3 & 17.5 & NHB   & 12.04.97 & 3600 &  IV &                             \nl
1506+498  & 15 06 53.10 & +49 52 13.5 & 17   & sd    & 12.03.97 & 3600 &  IV & CBS 302\tablenotemark{\dag} \nl
1506+496  & 15 06 19.60 & +49 37 11.4 & 16.0 & G     & 06.04.92 &  382 &  II & CSO 724\tablenotemark{\dag} \nl
1507+577  & 15 07 12.81 & +57 43 55.0 & 18.0 & Cont  & 18.03.86 &  780 &  II &                             \nl
1510+526  & 15 10 21.92 & +52 36 53.7 & 17.64& QSO   & 06.03.97 & 1800 &  IV &                             \nl
1510+586  & 15 10 57.56 & +58 36 07.0 & 17.0 & G     & 19.02.82 &  660 &  I  &                             \nl
1514+590  & 15 14 35.99 & +59 05 20.4 & 17.5 & DA    & 06.04.91 &  200 &  II &                             \nl
1516+494  & 15 16 54.27 & +49 29 53.9 & 17.5 & DA    & 14.04.97 & 3600 &  IV & CSO 737\tablenotemark{\dag} \nl
1516+519  & 15 16 14.18 & +51 55 40.0 & 17.0 & sd    & 06.04.91 &  250 &  II & CBS 310\tablenotemark{\dag} \nl
1517+553  & 15 17 27.36 & +55 22 44.5 & 17.5 & HBB   & 27.06.89 &  720 &  II &                             \nl
1520+545  & 15 20 50.64 & +54 33 32.5 & 17.5 & DA    & 06.04.91 &  246 &  II & CSO 745\tablenotemark{\dag} \nl
1522+545  & 15 22 35.48 & +54 33 29.4 & 17.0 & DA    & 27.06.89 &  720 &  II & CSO 750\tablenotemark{\dag} \nl
1522+551  & 15 22 48.83 & +55 11 22.4 & 17.5 & DA    & 06.04.91 &  333 &  II &                             \nl
1525+494  & 15 25 43.14 & +49 28 55.2 & 18.90& QSO   & 19.05.93 &  900 &  II & CSO 757\tablenotemark{\dag} \nl
1527+598  & 15 27 44.04 & +59 50 49.4 & 17.0 & HBB   & 05.04.92 &  720 &  II &                             \nl
1527+612B & 15 27 53.92 & +61 12 04.6 & 17.5 & DA    & 22.02.82 &  780 &  I  &                             \nl
1529+527  & 15 29 38.07 & +52 46 04.9 & 17.51& QSO   & 14.04.97 & 3600 &  IV & CSO 765\tablenotemark{\dag} \nl
1530+516  & 15 30 59.29 & +51 40 17.3 & 17   & NHB   & 11.04.96 &  600 &  IV &                             \nl
1532+547  & 15 32 52.91 & +54 43 43.6 & 16.0 & DA    & 26.06.89 &  480 &  II &                             \nl
1536+520  & 15 36 01.41 & +52 01 12.9 & 17.0 & DBA   & 05.04.92 &  540 &  II &                             \nl
1539+550  & 15 39 51.18 & +55 00 15.8 & 16.0 & sdB   & 26.06.89 &  480 &  II &                             \nl
1540+505  & 15 40 34.55 & +50 35 04.0 & 17.0 & DB    & 05.04.92 &  540 &  II &                             \nl
1541+495  & 15 41 49.10 & +49 32 04.9 & 17.0 & DA    & 05.04.92 &  540 &  II &                             \nl
1542+581  & 15 42 44.53 & +58 10 45.1 & 17.5 & Cont  & 11.04.81 &  720 &  I  &                             \nl
1542+517  & 15 42 51.33 & +51 42 33.6 & 17.0 & DA    & 05.04.92 &  660 &  II &                             \nl
1545+592  & 15 45 54.98 & +59 13 22.3 & 17.5 & G     & 19.05.93 &  240 &  II &                             \nl
1552+601  & 15 52 00.95 & +60 10 48.6 & 18.5 & F     & 12.02.91 &  840 &  II &                             \nl
1554+496B & 15 54 11.96 & +49 40 56.3 & 17.0 & NHB   & 06.04.92 &  720 &  II &                             \nl
1554+582  & 15 54 15.80 & +58 15 20.6 & 17.5 & sd    & 09.04.91 &  318 &  II &                             \nl
1556+605  & 15 56 01.24 & +60 32 51.1 & 17.0 & sdB   & 06.04.92 &  660 &  II &                             \nl
1600+587  & 16 00 23.68 & +58 42 57.5 & 15.0 & F     & 06.03.92 &  146 &  II &                             \nl
1600+575  & 16 00 52.32 & +57 35 39.8 & 18.5 & DA:   & 14.09.88 &  900 &  II &                             \nl
1604+606  & 16 04 53.70 & +60 41 42.7 & 17.0 & G     & 08.07.91 &  548 &  II &                             \nl
1612+554  & 16 12 09.64 & +55 28 58.2 & 16.5 & DA    & 26.06.89 &  540 &  II &                             \nl
1614+551  & 16 14 25.11 & +55 11 43.5 & 17.0 & sd:   & 26.06.89 &  660 &  II &                             \nl
1615+597  & 16 15 51.45 & +59 46 12.2 & 17.0 & G     & 10.07.91 &  472 &  II &                             \nl
1619+606  & 16 19 51.12 & +60 40 37.0 & 17.5 & DAF   & 22.02.82 &  720 &  I  &                             \nl
1621+564  & 16 21 36.19 & +56 29 38.3 & 17.0 & sdB   & 20.09.88 &  660 &  II &                             \nl
1621+558  & 16 21 47.30 & +55 51 12.2 & 17.0 & G     & 06.04.92 &  660 &  II &                             \nl
1622+587  & 16 22 05.88 & +58 47 30.0 & 17.5 & DB    & 14.05.85 &  780 &  II &                             \nl
1629+601  & 16 29 08.97 & +60 06 10.6 & 18.5 & DA:   & 10.04.81 &  900 &  I  &                             \nl
1629+590  & 16 29 38.75 & +59 04 28.7 & 17.5 & HBB   & 19.09.90 &  720 &  II &                             \nl
1642+515  & 16 42 16.23 & +51 30 28.6 & 16.5 & sdB-O & 06 03 92 &  260 &  II &                             \nl
1642+567  & 16 42 32.81 & +56 42 41.2 & 17.5 & DA    & 16.10.90 &  720 &  II &                             \nl
1643+582  & 16 43 38.59 & +58 12 48.9 & 18.0 & G     & 19.09.90 &  840 &  II &                             \nl
1655+588  & 16 55 36.83 & +58 52 44.9 & 16.5 & sdB   & 26.04.84 &  540 &  II &                             \nl
1655+591  & 16 55 44.41 & +59 09 21.8 & 16.5 & sd    & 17.04.97 & 3600 &  IV &                             \nl
1655+589  & 16 55 59.37 & +58 56 41.5 & 16.5 & HBB   & 15.04.97 & 3600 &  IV &                             \nl
1656+556  & 16 56 13.51 & +55 40 17.1 & 16.5 & sdO:  & 15.04.96 & 2400 &  IV &                             \nl
1657+584  & 16 57 52.25 & +58 28 00.0 & 17.5 & DA    & 26.09.90 &  780 &  II &                             \nl
1709+535  & 17 09 07.62 & +53 30 24.3 & 12.5 & DA:   & 05.04.90 &  120 &  II &                             \nl
1712+575  & 17 12 02.64 & +57 34 04.8 & 18.5 & G:    & 19.09.90 &  900 &  II &                             \nl
1712+578  & 17 12 20.86 & +57 53 31.2 & 17.0 & sdB   & 30.05.87 &  720 &  II &                             \nl
1712+593B & 17 12 25.42 & +59 23 00.8 & 17.5 & sdB   & 14.05.85 &  780 &  II &                             \nl
1712+493  & 17 12 32.60 & +49 19 34.3 & 14.5 &Pl.Neb.& 11.04.96 &  600 &  IV &                             \nl
1715+604  & 17 15 02.64 & +60 28 08.9 & 16.5 & NHB   & 26.04.87 &  660 &  II &                             \nl
1716+581  & 17 16 33.99 & +58 09 06.6 & 16.96& QSO   & 16.04.96 & 1200 &  IV &                             \nl
1717+583  & 17 17 03.13 & +58 18 19.3 & 17.38& QSO   & 16.04.96 & 1200 &  IV & RXJ1717.8+5815\tablenotemark{\ddag} \\
\enddata

\tablenotetext{\ddag}{~Bade et al. 1995.}
\tablenotetext{\S}{~Laurent-Muehleisen et al. 1998.}
\tablenotetext{\dag}{~Sanduleak \& Pesch 1989.}
\tablecomments{I -- SAO 6 m telescope (UAGS), II -- SAO 6 m telescope (IPCS),
III -- SAO 6 m telescope (LSS), IV -- GHO 2.1 m telescope (LFOSC).}
\end{planotable}

}

\clearpage

{\scriptsize{}
\begin{planotable}{cccccccccccccc}
\tablewidth{150mm}
\tablecaption{The results on the number of objects of different classes}
\tablehead{
\colhead{Type} & \colhead{QSO} & \colhead{WD} & \colhead{sd} &
\colhead{sdB} & \colhead{HBB} & \colhead{NHB} & \colhead{F} & \colhead{G} &
\colhead{Cont.} & \colhead{Comp.} & \colhead{CV} & \colhead{Pl.Neb.}
& \colhead{Pec.*}}
\startdata
Number & 35 &142 & 26 & 29 & 10 & 16 & 54 & 25 & 17 & 2 & 4 & 1 & 2 \nl
\enddata
\end{planotable}
}

\clearpage

{\scriptsize
\begin{planotable}{llllc|llll}
\tablewidth{165mm}
\tablecaption{Observed Emission Lines and Redshifts for the new QSOs}
\tablehead{
\colhead{Object} & \colhead{Redshift} & \colhead{$\lambda_{obs}$}  & \colhead{Identification} & &
\colhead{Object} & \colhead{Redshift} & \colhead{$\lambda_{obs}$}  & \colhead{Identification \vspace{2ex}}}
\startdata
0801+581  & 0.440  & 4029  & MgII~2798         & &  1201+517   &  0.803  & 5044  & MgII~2798                \nl
	  &        & 6255  & H$_{\gamma}$~4340 & &  1234+530   &  1.295  & 4381  & CIII]~1909               \nl
0806+505  & 1.205  & 6169  & MgII~2798         & &             &         & 6421  & MgII~2798                \nl
	  &        & 4209  & CIII]1909         & &  1251+572   &  1.273: & 6360  & MgII~2798                \nl
0839+541  & 0.216  & 3405  & MgII~2798         & &  1251+513   &  0.755: & 4910  & MgII~2798                \nl
	  &        & 4168  & [NeV] 3425        & &  1258+585   &  1.417  & 3745  & CIV~1549                 \nl
	  &        & 4535  & [OII] 3727         & &  	      &           & 4612  & CIII]~1909               \nl
	  &        & 4708  & [NeIII] 3869      & &  1305+538   &  0.803  & 5045  & MgII~2798                \nl
	  &        & 5281  & H$_{\gamma}$~4340 & &  1307+562   &  1.616  & 4053  & CIV~1549                 \nl
	  &        & 5916  & H$_{\beta}$~4861  & & 	      &           & 4992  & CIII]~1909               \nl
	  &        & 6035  & [OIII]~4959      & &  1315+605   &  1.981  & 4619  & CIV~1549                 \nl
	  &        & 6095  & [OIII]~5007       & &  	      &           & 5688  & CIII]~1909               \nl
0842+584  & 0.864  & 3555  & CIII]~1909      &   &  1338+551   &  1.637  & 4085  & CIV~1549                 \nl
	  &        & 5215  & MgII~2798        &  &  	      &           & 5030  & CIII]~1909               \nl
0916+513  & 0.545  & 4325  & MgII~2798       &  &  1345+592   &  0.768  & 4946  & MgII~2798                \nl
	  &        & 5977  & [NeIII] 3869    &   &  1417+510   &  1.415  & 4600  & CIII]~1909               \nl
	  &        & 6105  & [NeIII] 3968     & &  	      &           & 6765  & MgII~2798                \nl
0937+503  & 1.884  & 4061  & SiIV 1406        &  &  1417+514   &  1.305: & 3570  & CIV~1549                 \nl
	  &        & 4465  & CIV  1549        &  &             &         & 4400  & CIII]~1909               \nl
	  &        & 5047  & NIII] 1750      &  &  1423+500   &  2.220  & 2700  & Ly$_{\alpha}$/NV~1216    \nl
1016+510  & 1.314  & 4415  & CIII] 1909      &   &  	      &           & 4990  & CIV~1549                 \nl
	  & 	   & 6471  & MgII  2798       &  &  1424+502B  &  2.322  & 4040  & Ly$_{\alpha}$/NV~1216    \nl
1025+576  & 0.190  & 4705  & H$_{\epsilon}$~3970  & &   	      &           & 4651  & SiIV/OIV]~1400           \nl
	  &        & 4865  & H$_{\delta}$~4102  & &  	      &           & 5145  & CIV~1549                 \nl
	  &        & 5160  & H$_{\gamma}$~4340  & &  1435+500A &  1.550  & 3950  & CIV~1549                 \nl
	  &        & 5776  & H$_{\beta}$~4861   & &  	      &           & 4868  & CIII]~1909               \nl
	  &        & 5887  & [OIII]~4959        &  & 1447+552B &  0.366  & 5935  & H$_{\gamma}$~4340        \nl
	  &        & 5940  & [OIII]~5007        & &  	      &           & 6629  & H$_{\beta}$~4861         \nl
1047+557B & 0.331  & 3730  & MgII~2798          & &   1449+537  &  0.435  & 4012  & MgII~2798                \nl
	  &        & 5180  & H$_{8}$~3889       & &             &         & 6228  & H$_{\gamma}$~4340        \nl
	  &        & 5280  & H$_{\epsilon}$~3970& &             &         & 6975  & H$_{\beta}$~4861         \nl
	  &        & 5463  & H$_{\delta}$~4102  & &   1453+506  &  2.321: & 5145  & CIV~1549                 \nl
	  &        & 5783  & H$_{\gamma}$~4340  & &   	      &           & 6365  & CIII]~1909               \nl
	  &        & 6477  & H$_{\beta}$~4861   & &  1510+526  &  1.134  & 4073  & CIII] 1909               \nl
	  &        & 6602  & [OIII]~4959        & &             &         & 5970  & MgII  2798               \nl
	  &        & 6663  & [OIII]~5007        & &   1525+494  &  1.602  & 4030  & CIV 1549                 \nl
1058+561  & 0.935  & 3693  & CIII]~1909         & &             &         & 4966  & CIII] 1909               \nl
	  &        & 5415  & MgII~2798          & &   1529+527  &  1.360  & 4495  & CIII]~1909               \nl
1116+518  & 0.103  & 4523  & H$_{\delta}$~4102  & &             &         & 6618  & MgII~2798                \nl
	  &  	   & 4787  & H$_{\gamma}$~4340  & &   1716+581  &  0.580: & 4430  & MgII~2798                \nl
	  &        & 5362  & $H_{\beta}$~4861   & &   1717+583  &  0.311  & 5375  & $H_{\delta}$~4102        \nl
	  &        & 5470  & [OIII]~4959        & &            	&         & 5704  & $H_{\gamma}$~4340        \nl
	  &        & 5523  & [OIII]~5007        & &            	&         & 6000  & FeII~4570                \nl
1158+538  & 1.175  & 4152  & CIII] 1909         & &            	&         & 6380  & $H_{\beta}$~4861         \nl
	  &        & 6085  & MgII 2798          & &            	&         & 6560  & [OIII]~5007              \nl
\enddata
\end{planotable}
}


\begin{references}

\reference{} Afanasiev, V.L., Lipovetsky, V.A., Mikhailov, V.P, Nazarov, E.A.,
\& Shapovalova, A.I. 1991, Astrofiz. Issled. (Izv. SAO.), 31, 121

\reference{} Afanasiev, V.L., Burenkov, A.N., Vlasyuk, V.V., \& Drabek, S.V.
1995, SAO technical report No. 234

\reference{} Bade, N., Fink, H.H., Engels, D., Voges, W. Hagen, H.-J., Wisotzki, L., 
\& Reimers, D. 1995, A\&AS, 110, 469

\reference{} Berg, C., Wegner, G., Foltz, C.B., Chaffee, F.H., \& Hewett,
P.C. 1992, ApJS, 78, 409

\reference{} Bicay, M.D, Seal, J., Kojoian, G., Stepanian, J.A.,
Chavushyan, V.H., Erastova, L.K., \& Ayvazyan V.T. 1999, (in press)

\reference{} Chavushyan, V.H., Stepanian, J.A., Balayan, S.K.,\& Vlasyuk,
V.V. 1995, Astron. Letters. 21, 894

\reference{} Chavushyan, V.H., Stepanian, J.A., Vlasyuk, V.V., \& Ayvazyan, V.T.
 1999 (in preparetion)

\reference{} Drabek, S.V., Kopilov, I.M., Somov, N.N., \& Somova, T.A.
1985, Astrofiz. Issled. (Izv. SAO.), 22, 64

\reference{} Green, R.F., Schmidt, M., \& Liebert, J. 1986, ApJS, 61, 305

\reference{} Laurent-Muehleisen, S.A., Kollgaard, R.I., Ciardullo, R., Feigelson, E.D., Brinkmann, W., \& Siebert, J. 1998, ApJS, 118, 127

\reference{} Markarian, B.E., \& Stepanian, J.A. 1983, Astrofizika, 19, 639

\reference{} Markarian, B.E., Stepanian, J.A., \& Lipovetsky, V.A.  1980,
 Astron. Zirk., 1141, 1

\reference{} Markarian, B.E., Stepanian, J.A., \& Lipovetsky, V.A.  1980,
 Astron. Zirk., 1142, 1

\reference{} Markarian, B.E., Stepanian, J.A., \& Lipovetsky, V.A. 1982,
 Astron. Zirk., 1233, 2

\reference{} Markarian, B.E., Stepanian, J.A., \& Lipovetsky, V.A.  1983,
 Astron. Zirk., 1265, 1

\reference{} Markarian, B.E., Stepanian, J.A., \& Lipovetsky, V.A.,  1984,
 Astron. Zirk., 1346, 1

\reference{} Markarian, B.E., Stepanian, J.A., \& Lipovetsky, V.A.  1985,
 Astron. Zirk., 1381, 1

\reference{} Markarian, B.E., Erastova L.K., Stepanian, J.A.,
Lipovetsky V.A., \& Shapovalova A.I. 1987, Pis'ma v AZh, 13, 1

\reference{} Markarian, B.E., Lipovetsky V.A., \& Stepanian, J.A., 1987,
Astrofizica, 19, 29

\reference{} Sanduleak, N., \& Pesch, P. 1989, ApJS, 70, 173

\reference{} Sion, E.M., Greenstein, J.L., Landstreet, J., Shipman, H.L.,
\& Wegner, G. 1983, ApJ, 269, 253

\reference{} Stepanian, J.A., Lipovetsky, V.A., \& Erastova, L.K. 1990a,
Astrofizika, 32, 441

\reference{} Stepanian, J.A., Lipovetsky, V.A., Shapovalova, A.I.,
\& Erastova, L.K. 1990b, Astrofizika, 33, 89

 \reference{} Stepanian, J.A., Lipovetsky, V.A., Shapovalova, A.I.,
Erastova, L.K., \& Chavushyan, V.H. 1990c, Astrofizika, 33, 199

\reference{} Stepanian, J.A., Lipovetsky, V.A., Shapovalova, A.I., Erastova,
L.K., \& Chavushyan, V.H. 1990d, Astrofizika, 33, 351

\reference{} Stepanian, J.A., Lipovetsky, V.A., Chavushyan, V.H., Erastova,
L.K., \& Shapovalova, A.I. 1991a, Astrofizika, 34, 5

\reference{} Stepanian, J.A., Lipovetsky, V.A., Chavushyan, V.H., Erastova,
L.K., \& Shapovalova, A.I. 1991b, Astrofizika, 34, 315

\reference{} Stepanian, J.A., Lipovetsky, V.A., Chavushyan, V.H., Erastova,
L.K., \& Balayan, S.K. 1993, Bull. Spec. Astrophys Obs. (Izv. SAO), 36, 5

\reference{} Stepanian, J.A. 1994, D.Sci. Thesis , Nizhnij Arkhys

\reference{} Vlasyuk, V.V. 1993, Bull. Spec. Astrophys. Obs. (Izv. SAO.),
36, 107
\reference{} Zickgraf, F.J., Thiering, I., Krautter, J. et al. 1997, A\&AS,123,103
\end{references}
\end{document}